\documentclass[conference, a4paper]{IEEEtran}

\usepackage{setspace, dsfont, enumerate}
\usepackage{amssymb, amsmath, mathrsfs, amsthm, stmaryrd, upgreek, mathtools}
\usepackage{pgfplots, caption, tikz, subfigure, graphicx}
\usepackage[scr=boondoxo]{mathalfa}
\usepackage{url, hyperref, cite}

\AtBeginDocument{
  \addtolength\abovedisplayskip{-2.2pt}
  \addtolength\belowdisplayskip{-2.2pt}
}

\input{def.tex}

\begin{document}

\title{Deterministic Performance Analysis of Subspace Methods for Cisoid Parameter Estimation}

\author{\IEEEauthorblockN{C\'eline Aubel and Helmut B\"olcskei \medskip}
\IEEEauthorblockA{Dept.~IT~\&~EE, ETH Zurich, Switzerland\\
Email: \{aubelc, boelcskei\}@nari.ee.ethz.ch}
}

\maketitle

\linespread{0.96}



\begin{abstract} \boldmath
	Performance analyses of subspace algorithms for cisoid parameter estimation available in the literature are predominantly of statistical nature with a focus on asymptotic---either in the sample size or the SNR---statements. This paper presents a deterministic, finite sample size, and finite--SNR performance analysis of the ESPRIT algorithm and the matrix pencil method. Our results are based, inter alia, on a new upper bound on the condition number of Vandermonde matrices with nodes inside the unit disk. This bound is obtained through a generalization of Hilbert's inequality frequently used in large sieve theory.
\end{abstract}

\vspace{-0.15cm}
\section{Introduction}

\newcommand{\fromto}[1]{= #1}

The foundation of high-resolution methods for estimating the parameters of a sum of complex exponentials was laid by Prony~\cite{Prony1795} and refined by Pisarenko~\cite{Pisarenko1973}. Both Prony's and Pisarenko's method are very sensitive to additive noise~\cite{Stoica1988}. Modern high-resolution estimation methods relying on subspace concepts exhibit less noise sensitivity. Prominent subspace methods are the MUltiple SIgnal Classification (MUSIC) algorithm~\cite{Schmidt1986}, the Estimation of Signal Parameters via Rotational Invariance Techniques (ESPRIT)~\cite{Roy1986} algorithm, and the Toeplitz Approximation Method (TAM)\cite{Kung1983}. Originally developed for undamped sinusoids, all of the above techniques were later found to also be applicable to exponentially damped sinusoids. Other subspace methods, specifically designed for exponentially damped sinusoids, include the Kumaresan-Tufts (KT) algorithm~\cite{Kumaresan1982}, and the Matrix Pencil (MP) method~\cite{Hua1990}. A survey of subspace estimation methods can be found in \cite{Stoica2005}. 

The problem of estimating the parameters of a sum of damped or undamped sinusoids arises in numerous practical applications such as 
direction finding in array processing~\cite{Schelle1993}, velocity and acceleration estimation from Lidar or Radar echoes~\cite{Besson1993, Chen1995}, super-resolution~\cite{Candes2014}, sampling of signals with finite rate of innovation~\cite{Vetterli2002}, line spectral estimation~\cite{Tang2013}, spectrum analysis of musical signals~\cite{Laroche1993}, and speech signal analysis and synthesis \cite{McAulay1986}. 

Formally, the problem considered in this paper is as follows. Recover the complex numbers $\node_1, \node_2, \ldots, \node_\nbNodes$,  with $\abs{\node_k} \leq 1$, $k \fromto{1, 2, \ldots, \nbNodes}$, henceforth referred to as ``nodes'', and the corresponding complex weights $\alpha_1, \alpha_2, \ldots, \alpha_\nbNodes$ from the noisy measurements $\noisySig_n \triangleq \sig_n + e_n$, $n \fromto{0, 1, \ldots, \nbSamples-1}$, where
\vspace{-0.15cm}
\begin{equation}
	x_n \triangleq \sum_{k = 1}^\nbNodes \alpha_k z_k^n,
	\label{eq: sum of exponentially damped sinusoids} 
	\\[-0.15cm]
\end{equation}
$e_n$ is deterministic noise, and the number of samples $\nbSamples$ satisfies $\nbSamples \geq 2\nbNodes$. The complex numbers $\node_1, \node_2, \ldots, \node_\nbNodes$ can be written as $\node_k = e^{-d_k}e^{2\pi if_k/F_\mathrm{s}}$, $k \fromto{1, 2, \ldots, \nbNodes}$, where $d_k \geq 0$ is the damping factor and $f_k$ the frequency of the $k$-th sinusoid, and $F_\mathrm{s}$ is the sampling frequency corresponding to the number of samples taken per unit time.

There is a vast literature on statistical performance analysis of subspace methods~\cite{Rao1988,Hua1990,Hua1991,Li1991,Eriksson1993,Yuen1996}. 
The setup in this line of work is to take the parameters $\alpha_k$ and $\node_k$ as random and to analyze the bias and the statistical efficiency of various estimators in the large $\nbSamples$ and/or high signal-to-noise ratio ($\mathrm{SNR}$) limit.
Deterministic (with respect to the parameters to be estimated and to additive noise), non-asymptotic performance results became available only recently for the MUSIC algorithm in \cite{Liao2014} and for a new variant of the MP method in \cite{Moitra2014}. Both \cite{Liao2014} and \cite{Moitra2014} apply, however, to undamped sinusoids only, i.e., $\abs{\node_k} = 1$, $k \fromto{1, 2, \ldots, \nbNodes}$. 
The main statements in \cite{Liao2014, Moitra2014} are based on new upper bounds on the condition number of $L \times \nbNodes$ ($L \geq \nbNodes$) Vandermonde matrices with nodes $\node_1, \node_2, \ldots, \node_\nbNodes$ on the unit circle (i.e., $\abs{\node_k} = 1$, for all $k \fromto{1, 2, \ldots, \nbNodes}$). To the best of our knowledge, no deterministic performance analysis exists for the ESPRIT algorithm.

\textit{Contributions.} In this paper, we present a deterministic, finite--$\nbSamples$, and finite--$\mathrm{SNR}$ performance analysis of the ESPRIT algorithm and the classical MP method.
Our results apply to both undamped and damped sinusoids, i.e., $\abs{\node_k} \leq 1$, $k \fromto{1, 2, \ldots, \nbNodes}$.
A central technical element of our proofs is a new upper bound on the condition number of Vandermonde matrices with nodes $\node_1, \node_2, \ldots, \node_\nbNodes$ in the complex unit disk (i.e., $\abs{\node_k} \leq 1$, for $k \fromto{1, 2, \ldots, \nbNodes}$). This bound is established through a generalization of Hilbert's inequality~\cite{Montgomery1998} and shows that the condition number remains close to $1$ if the minimum wrap-around distance between the node frequencies $f_k$ 
is large relative to $F_\mathrm{s}/(\nbSamples-1)$, and if the nodes $z_k$ remain close to the unit circle. Throughout the paper proofs are omitted due to space constraints.

\textit{Notation.}
The complex conjugate of $z \in \C$ is~$\overline{z}$. Lowercase boldface letters stand for column vectors and uppercase boldface letters denote matrices. The superscripts $^T$ and $^H$ designate transposition and Hermitian transposition, respectively. For a vector $\boldsymbol{x} \triangleq \{x_k\}_{k = 1}^K \in \C^K$, we write $\norm{\boldsymbol{x}}_2$ for its $\ell^2$-norm, that is, $\norm{\boldsymbol{x}}_2 \triangleq \left(\sum_{k = 1}^K \abs{x_k}^2\right)^{1/2}$.
We denote the smallest and largest singular value of $\mathbf{A} \in \C^{M \times N}$  by $\sigma_\mathrm{min}(\mathbf{A})$ and $\sigma_\mathrm{max}(\mathbf{A})$, respectively. The condition number of $\mathbf{A} \in \C^{M \times N}$ is $\kappa(\mathbf{A}) \triangleq \sigma_\mathrm{max}(\mathbf{A})/\sigma_\mathrm{min}(\mathbf{A})$.
The generalized eigenvalues of the pair $(\dataMatrixOne, \dataMatrixTwo)$, with $\dataMatrixOne, \dataMatrixTwo \in \C^{L \times L}$, are the values of $\lambda$ for which there exists $\boldsymbol{y} \neq \boldsymbol{0}$ with $\dataMatrixTwo\boldsymbol{y} = \lambda\dataMatrixOne\boldsymbol{y}$.
For $L \in \N$ such that $L \geq \nbNodes$, we define the Vandermonde matrix
\begin{equation*}
	\vander_L \triangleq \begin{pmatrix} 1 & 1 & \ldots & 1 & 1\\
			\node_1 & \node_2 & \ldots & \node_{\nbNodes-1} & \node_\nbNodes \\
			\node_1^2 & \node_2^2 & \ldots & \node_{\nbNodes-1}^2 & \node_\nbNodes^2 \\
			\vdots & \vdots & \ddots & \vdots & \vdots \\
			\node_1^{L-1} & \node_2^{L-1} & \ldots & \node_{\nbNodes-1}^{L-1} & \node_\nbNodes^{L-1}
		\end{pmatrix} 
	\in \C^{L \times \nbNodes},
\end{equation*}
where the $\node_1, \node_2, \ldots, \node_\nbNodes$ are the nodes in \eqref{eq: sum of exponentially damped sinusoids}.
$\diag(a_1, a_2, \ldots, a_L) \in \C^{L \times L}$ denotes the diagonal matrix with $a_1, a_2, \ldots, a_L$ on its main diagonal.
For complex numbers $x_0, x_1, \ldots, x_{N-1}$ and $L \in \N$ with $1 \leq L \leq N$, $\hankel{L}{x_0, x_1, \ldots, x_{N-1}}$ designates the (rectangular) Hankel matrix
\vspace{-0.1cm}
\begin{align*}
	&\mathscr{H}_L(x_0, x_1, \ldots, x_{N-1}) \triangleq \\
		&\hspace{-0.2cm} \begin{pmatrix} x_0 & x_1 & \cdots & x_{N-L-1} & x_{N-L} \\
		x_1 & x_2 & \cdots & x_{N-L} & x_{N-L+1} \\
		\vdots & \vdots & \ddots & \vdots & \vdots \\
		x_{L-2} & x_{L-1} & \cdots & x_{N-3} & x_{N-2} \\
		x_{L-1} & x_{L} & \cdots & x_{N-2} & x_{N-1} \end{pmatrix} \in C^{L \times (N-L+1)}.
\end{align*}


\section{Subspace methods}

Before stating our main results in Section~\ref{sec: stability analysis}, we summarize the ESPRIT algorithm and the MP method.
In the remainder of the paper, we assume that the nodes $\node_1, \node_2, \ldots, \node_\nbNodes$ are non-zero and pairwise distinct, i.e., $\node_{k_1} \neq \node_{k_2}$ for $k_1 \neq k_2$. We furthermore take, throughout, $\nbSamples \geq 2\nbNodes$ and let $\sizeData \in \N$ such that $\nbNodes \leq L \leq \nbSamples-\nbNodes$.

\vspace{-0.05cm}
\subsection{ESPRIT algorithm}
\vspace{-0.08cm}

We start by constructing the data matrix $\noisyDataMatrix \triangleq \hankel{\sizeData}{\noisySig_0, \noisySig_1, \ldots, \noisySig_{\nbSamples-1}} \in \C^{\sizeData \times (\nbSamples-\sizeData+1)}$, which satisfies $\noisyDataMatrix = \dataMatrix + \noiseMatrix$,
where $\dataMatrix \triangleq \hankel{\sizeData}{\sig_0, \sig_1, \ldots, \sig_{\nbSamples-1}} \in \C^{\sizeData \times (\nbSamples-\sizeData+1)}$ and $\noiseMatrix \triangleq \hankel{\sizeData}{\noise_0, \noise_1, \ldots, \noise_{\nbSamples-1}} \in \C^{\sizeData \times (\nbSamples-\sizeData+1)}$, $\noisySig_n = \sig_n + e_n$ with $\sig_n$ as in \eqref{eq: sum of exponentially damped sinusoids} and $\noise_n$ deterministic noise. In the noiseless case, $\dataMatrix$ can be factorized according to $\dataMatrix = \vander_\sizeData\mathbf{D}\vander_{\nbSamples-\sizeData+1}^T$, it has $\nbNodes$ non-zero singular values $\lambda_1, \lambda_2, \ldots, \lambda_\nbNodes$, and can be decomposed as
\vspace{-0.1cm}
\begin{equation}
	\dataMatrix \triangleq \underbrace{\begin{pmatrix} \subspaceMatrix & \noiseSubspaceMatrix \end{pmatrix}}_{\reversetriangleq \mathbf{U}} \begin{pmatrix}\boldsymbol{\Lambda} & \boldsymbol{0} \\ \boldsymbol{0} & \boldsymbol{0} \end{pmatrix} \underbrace{\begin{pmatrix} \mathbf{R}^H \\ \mathbf{R}_\perp^H\end{pmatrix}}_{\reversetriangleq \mathbf{W}^H} = \subspaceMatrix \boldsymbol{\Lambda} \mathbf{R}^H,
	\label{eq: data matrix SVD} 
\end{equation}
where $\mathbf{U} \in \C^{\sizeData \times \sizeData}$ and $\mathbf{W} \in \C^{(\nbSamples-\sizeData+1) \times (\nbSamples-\sizeData+1)}$ are unitary, and $\boldsymbol{\Lambda} \triangleq \diag(\lambda_1, \lambda_2, \ldots, \lambda_\nbNodes) \in \R^{\nbNodes \times \nbNodes}$. 

The ESPRIT algorithm relies on the following rotational invariance property of the subspace $\subspace$ spanned by the columns of $\vander_L$. 
Let $\vander_\downarrow \in \C^{(\sizeData-1) \times \nbNodes}$ be the matrix consisting of the $\sizeData-1$ first rows of $\vander_L$ and $\vander_\uparrow \in \C^{(\sizeData-1) \times \nbNodes}$ the matrix consisting of the $\sizeData-1$ last rows of $\vander_L$. We have
\begin{equation*}
	\vander_\uparrow = \vander_\downarrow\mathbf{J}, \qquad \text{where} \quad \mathbf{J} \triangleq \diag(\node_1, \node_2, \ldots, \node_K).
\end{equation*}
Since the columns of both $\subspaceMatrix$ and $\vander_L$ are bases for $\subspace$, there exists an invertible matrix $\mathbf{P} \in \C^{\nbNodes \times \nbNodes}$ such that $\subspaceMatrix = \vander_L\mathbf{P}$.
Next, letting $\mathbf{S}_\downarrow \in \C^{(\sizeData-1) \times \nbNodes}$ denote the matrix consisting of the $\sizeData-1$ first rows of $\mathbf{S}$ and $\mathbf{S}_\uparrow \in \C^{(\sizeData-1) \times \nbNodes}$ the matrix consisting of the $\sizeData-1$ last rows of $\mathbf{S}$, it follows from $\vander_\uparrow = \vander_\downarrow\mathbf{J}$ that $\mathbf{S}_\uparrow = \mathbf{S}_\downarrow\boldsymbol{\Phi}$, where $\boldsymbol{\Phi} \triangleq \mathbf{P}^{-1}\mathbf{J}\mathbf{P}$. As $\mathbf{J} \triangleq \diag(\node_1, \node_2, \ldots, \node_K)$ and $\mathbf{P} \in \C^{\nbNodes \times \nbNodes}$ is invertible, $z_1, z_2, \ldots, z_K$ are the eigenvalues of the matrix $\boldsymbol{\Phi}$. 

In the noisy case, we have to work on $\noisyDataMatrix$ (instead of $\dataMatrix$), which does not have rank $\nbNodes$, and might actually even be of full rank. The basic idea here is to identify the signal and noise subspaces, and to split the measurements into corresponding sets. This can be done by decomposing $\noisyDataMatrix$ along the lines of \eqref{eq: data matrix SVD} to get
\vspace{-0.2cm}
\begin{equation}
	\noisyDataMatrix = \underbrace{\begin{pmatrix} \noisySubspaceMatrix & \noisyNoiseSubspaceMatrix \end{pmatrix}}_{\reversetriangleq \widetilde{\mathbf{U}}} \begin{pmatrix}\boldsymbol{\widetilde{\Lambda}} & \boldsymbol{0} \\ \boldsymbol{0} & \widetilde{\boldsymbol{\Gamma}} \end{pmatrix} \underbrace{\begin{pmatrix} \widetilde{\mathbf{R}}^H \\ \widetilde{\mathbf{R}}_\perp^H\end{pmatrix}}_{\reversetriangleq \widetilde{\mathbf{W}}^H},
	\label{eq: noisy SVD of the noisy data matrix} 
\end{equation}
where $\widetilde{\mathbf{U}} \in \C^{\sizeData \times \sizeData}$ and $\widetilde{\mathbf{W}} \in \C^{(\nbSamples-\sizeData+1) \times (\nbSamples-\sizeData+1)}$ are unitary, $\widetilde{\boldsymbol{\Lambda}} \in \R^{\nbNodes \times \nbNodes}$ is a diagonal matrix containing the $\nbNodes$ largest singular values of $\noisyDataMatrix$, and $\widetilde{\boldsymbol{\Gamma}} \in \R^{(\sizeData-\nbNodes) \times (\nbSamples-\sizeData-\nbNodes+1)}$ is a rectangular diagonal matrix containing the remaining singular values of $\noisyDataMatrix$. In the noisy case, the ESPRIT algorithm then proceeds by applying the procedure outlined above to $\noisySubspaceMatrix$ instead of $\subspaceMatrix$: the estimates $\widehat{z}_1, \widehat{z}_2, \ldots, \widehat{z}_K$ are thus given by the eigenvalues of the matrix $\widetilde{\boldsymbol{\Phi}} = \widetilde{\mathbf{S}}_\downarrow^\dagger\widetilde{\mathbf{S}}_\uparrow \in \C^{K \times K}$. 
Formally, we write $\widehat{\boldsymbol{\node}} = \text{\textsc{Esprit}}(\noisySigVector, \nbNodes, \sizeData)$ for the estimates $\widehat{\boldsymbol{\node}} \triangleq \left(\widehat{\node_1}\ \widehat{\node}_2\ \ldots \ \widehat{\node}_\nbNodes\right)^T \in \C^\nbNodes$ delivered by the ESPRIT algorithm. Note that throughout the paper, we consider the least-squares (LS)-ESPRIT algorithm as introduced in \cite{Roy1986}.

\vspace{-0.1cm}
\subsection{MP method}

We start by building the data matrices
\begin{align*}
	\noisyDataMatrixOne&\triangleq \hankel{L}{\widetilde{x}_0, \widetilde{x}_1, \ldots, \widetilde{x}_{\nbSamples-3}, \widetilde{x}_{\nbSamples-2}} \in \C^{L \times (\nbSamples-L)}\\
	\noisyDataMatrixTwo&\triangleq \hankel{L}{\widetilde{x}_1, \widetilde{x}_2, \ldots, \widetilde{x}_{\nbSamples-2}, \widetilde{x}_{\nbSamples-1}} \in \C^{L \times (\nbSamples-L)},
\end{align*}
and noting that $\noisyDataMatrixOne= \dataMatrixOne+ \noiseMatrixOne$ and $\noisyDataMatrixTwo= \dataMatrixTwo+ \noiseMatrixTwo$, where
\begin{align}
	\dataMatrixOne &\triangleq \hankel{L}{x_0, x_1, \ldots, x_{\nbSamples-3}, x_{\nbSamples-2}} \label{eq: expression X1 MP method} \\ 
	\dataMatrixTwo &\triangleq \hankel{L}{x_1, x_2, \ldots, x_{\nbSamples-2}, x_{\nbSamples-1}} \label{eq: expression X2 MP method} \\ 
	\noiseMatrixOne &\triangleq \hankel{L}{e_0, e_2, \ldots, e_{\nbSamples-3}, e_{\nbSamples-2}} \notag \\
	\noiseMatrixTwo &\triangleq \hankel{L}{e_1, e_2, \ldots, e_{\nbSamples-2}, e_{\nbSamples-1}}. \notag
\end{align}
The MP method relies on the fact that in the noiseless case, the matrices $\dataMatrixOne$ and $\dataMatrixTwo$ can be factorized according to $\dataMatrixOne =\vander_L\mathbf{D}_{\boldsymbol{\alpha}}\vander_{\nbSamples-L}^T$ and $\dataMatrixTwo = \vander_L\mathbf{D}_{\boldsymbol{\alpha}}\mathbf{D}_{\boldsymbol{\node}}\vander_{\nbSamples-L}^T$, where $\mathbf{D}_{\boldsymbol{\alpha}} \triangleq \diag(\alpha_1, \alpha_2, \ldots, \alpha_\nbNodes)$ and $\mathbf{D}_{\boldsymbol{\node}} \triangleq \diag(\node_1, \node_2, \ldots, \node_\nbNodes)$. This factorization implies that the nodes $\node_1, \node_2, \ldots, \node_\nbNodes$ are specified uniquely by the non-zero values of $\lambda$ for which the rank of the matrix pencil $\dataMatrixTwo - \lambda\dataMatrixOne$ drops by one relative to the rank of the pencil for all other values of $\lambda$.

In the noisy case, $\dataMatrixOne$ and $\dataMatrixTwo$ are replaced by $\noisyDataMatrixOne$ and $\noisyDataMatrixTwo$ and $\dataMatrixTwo - \lambda\dataMatrixOne$ by the associated pencil $\noisyDataMatrixTwo - \lambda\noisyDataMatrixOne$. It will, in general, no longer be possible to extract the nodes by determining the rank-reducing values of $\lambda$. Instead, we define
\begin{align}
	\widetilde{\boldsymbol{\Psi}}_1 &\triangleq \widetilde{\subspaceMatrix}_1^H\noisyDataMatrixOne\widetilde{\mathbf{R}}_1 \in \C^{\nbNodes \times \nbNodes} \label{eq: definition psi 1}\\ 
	\widetilde{\boldsymbol{\Psi}}_2 &\triangleq \widetilde{\subspaceMatrix}_1^H\noisyDataMatrixTwo\widetilde{\mathbf{R}}_1 \in \C^{\nbNodes \times \nbNodes}, \label{eq: definition psi 2} 
\end{align}
where $\widetilde{\subspaceMatrix}_1 \in \C^{L \times \nbNodes}$ and $\widetilde{\mathbf{R}}_1 \in \C^{(\nbSamples-L) \times \nbNodes}$ are obtained through the singular value decomposition
\begin{equation*}
	 \noisyDataMatrixOne = \widetilde{\mathbf{U}}_1\widetilde{\boldsymbol{\Sigma}}\widetilde{\mathbf{W}}_1^H = \begin{pmatrix} \noisySubspaceMatrix_1 & \noisySubspaceMatrix_{1, \perp} \end{pmatrix}\begin{pmatrix} \widetilde{\boldsymbol{\Lambda}} & \boldsymbol{0} \\ \boldsymbol{0} & \widetilde{\boldsymbol{\Gamma}}\end{pmatrix} \begin{pmatrix} \widetilde{\mathbf{R}}_1^H \\ \widetilde{\mathbf{R}}_{1,\perp}^H\end{pmatrix},
\end{equation*}
and $\widetilde{\boldsymbol{\Lambda}} \in \C^{\nbNodes \times \nbNodes}$ contains the $\nbNodes$ largest singular values of $\noisyDataMatrixOne$. Again, this singular value decomposition extracts the signal and noise subspaces. The matrices $\widetilde{\boldsymbol{\Psi}}_1$ and $\widetilde{\boldsymbol{\Psi}}_2$ are constructed from the signal subspace, and the MP method estimates the nodes by identifying the generalized eigenvalues (counted with their algebraic multiplicities) of $(\widetilde{\boldsymbol{\Psi}}_1, \widetilde{\boldsymbol{\Psi}}_2)$, that we denote by $\widehat{z}_1, \widehat{z}_2, \ldots, \widehat{z}_K$. In the noiseless case, the resulting estimates are equal to the true nodes $\node_1, \node_2, \ldots, \node_\nbNodes$. In the noisy case, $\widetilde{\boldsymbol{\Psi}}_1$ and $\widetilde{\boldsymbol{\Psi}}_2$ may be singular. If this is, indeed, the case, the polynomial $\widetilde{P}(\lambda) \triangleq \det(\widetilde{\boldsymbol{\Psi}}_2 - \lambda \widetilde{\boldsymbol{\Psi}}_1)$ has fewer than $\nbNodes$ roots, say $Q \leq \nbNodes$, and hence, $(\widetilde{\boldsymbol{\Psi}}_1, \widetilde{\boldsymbol{\Psi}}_2)$ has $Q \leq \nbNodes$ generalized eigenvalues. The ``missing'' $\nbNodes - Q$ values can then be thought of as generalized eigenvalues that are infinite in the sense that vectors $\boldsymbol{y} \neq \boldsymbol{0}$ in the null-space of $\widetilde{\boldsymbol{\Psi}}_1$ (i.e., $\widetilde{\boldsymbol{\Psi}}_1\boldsymbol{y} = 0\widetilde{\boldsymbol{\Psi}}_2\boldsymbol{y}$) are generalized eigenvectors of $(\widetilde{\boldsymbol{\Psi}}_2, \widetilde{\boldsymbol{\Psi}}_1)$ corresponding to the generalized eigenvalue $\lambda^{-1} = 0$, and hence $\lambda = \infty$. In the remainder of the paper, we therefore extend the complex plane by adding a point denoted by $\infty$ and assigned to the estimated nodes that correspond to $\boldsymbol{y} \in \C^\nbNodes \setminus \{\boldsymbol{0}\}$ satisfying $\widetilde{\boldsymbol{\Psi}}_1\boldsymbol{y} = \boldsymbol{0}$.
Throughout, $\widehat{\boldsymbol{\node}} = \text{\textsc{Mp}}(\noisySigVector, \nbNodes, \sizeData)$ refers to the estimates $\widehat{\boldsymbol{\node}} \triangleq \left(\widehat{\node_1}\ \widehat{\node}_2\ \ldots \ \widehat{\node}_\nbNodes\right)^T \in (\C\cup\{\infty\})^\nbNodes$ delivered by the MP method corresponding to the inputs $(\noisySigVector, \nbNodes, \sizeData)$.


\section{Performance analysis of subspace methods}
\label{sec: stability analysis} 

The statistical performance results in \cite{Rao1988,Stoica1989,Hua1990,Hua1991,Li1991,Eriksson1993,Yuen1996} assume that the parameters $\alpha_k$ and $\node_k$ and noise $\noise_n$ are all random, and quantify the bias and the statistical efficiency of various estimators. Analytical expressions are typically, however, possible only in the asymptotic regimes $\nbSamples \rightarrow \infty$ and/or $\mathrm{SNR} \rightarrow \infty$. In this section, we provide a deterministic, finite--$\nbSamples$, and finite--$\mathrm{SNR}$ performance analysis of the ESPRIT algorithm and the MP method. The corresponding results apply to bounded, but otherwise arbitrary, deterministic noise and assume the model order $\nbNodes$ to be known.

We want to quantify the Euclidean distance between the estimated nodes $\widehat{\node}_1, \widehat{\node}_2, \ldots, \widehat{\node}_\nbNodes$ and the true nodes $\node_1, \node_2, \ldots, \node_\nbNodes$. 
We will also need the chordal distance between points of the extended complex plane $\C \cup \{\infty\}$.

\begin{defn}[Chordal distance]
	The chordal distance between $z \in (\C \cup \{\infty\})$ and $z' \in (\C \cup \{\infty\})$ is defined as
	\begin{equation*}
		\chi(z, z') \triangleq \begin{cases}\displaystyle\frac{\abs{z - z'}}{\sqrt{1 + \abs{z}^2} \sqrt{1 + \abs{z'}^2}}, & z, z' \in \C\\
			\displaystyle\frac{1}{\sqrt{1 + \abs{z}^2}}, & z \in \C, z' = \infty.
			\end{cases}
	\end{equation*}
\end{defn}

Furthermore, we will need the concept of regular pairs of matrices.

\begin{defn}[Regular pair]
	Let $\mathbf{A}, \mathbf{B} \in \C^{\nbNodes \times \nbNodes}$. The pair of matrices $(\mathbf{A}, \mathbf{B})$ is said to be regular if and only if there exists $(\alpha, \beta) \in \C^2$ such that $\det(\alpha\mathbf{A} - \beta\mathbf{B}) \neq 0$.
\end{defn}

It is shown in \cite[Chap.~VI]{Stewart1990} that regular pairs of matrices $(\mathbf{A}, \mathbf{B})$, with $\mathbf{A}, \mathbf{B} \in \C^{\nbNodes \times \nbNodes}$, have $\nbNodes$ generalized eigenvalues in $\C\cup\{\infty\}$ (counted with their algebraic multiplicities).

We are now ready to present our main result.

\begin{thm}
	\label{thm: performance analysis for both MP and LS-ESPRIT} 
	For $k \fromto{1, 2, \ldots, \nbNodes}$, let $\node_k$ be complex nodes in the unit disk, i.e., $\abs{\node_k} \leq 1$. Let $\widehat{\boldsymbol{\node}} \triangleq \left(\widehat{\node_1}\ \widehat{\node}_2\ \ldots \ \widehat{\node}_\nbNodes\right)^T \in (\C\cup\{\infty\})^\nbNodes$ be given by $\widehat{\boldsymbol{\node}} = \text{\textsc{Alg}}(\noisySigVector, \nbNodes, \sizeData)$, where $\noisySigVector \triangleq \left(\noisySig_0\ \noisySig_1\ \ldots \ \noisySig_{\nbSamples-1}\right)^T \in \C^\nbSamples$ is the measurement vector defined by 
	\vspace{-0.1cm}
	\begin{equation*}
		\noisySig_n \triangleq \sig_n + \noise_n, \quad \text{with }\quad \sig_n \triangleq \sum_{k = 1}^\nbNodes \alpha_k \node_k^n, \\[-0.1cm]
	\end{equation*}
	$\noiseVector \triangleq \left(\noise_0\ \noise_1\ \ldots\ \noise_{\nbSamples-1}\right) \in \C^\nbSamples$ is a bounded noise term, $\nbNodes$ is the number of nodes (assumed known) to be recovered, and $\sizeData$ is an integer such that $\nbNodes \leq L \leq \nbSamples-\nbNodes$. Furthermore, we let $\nbSamples \geq 2\nbNodes$, $\alpha_\mathrm{min} \triangleq \displaystyle \min_{1\leq k \leq \nbNodes} \abs{\alpha_k}$, and $\alpha_\mathrm{max} \triangleq \displaystyle \max_{1\leq k \leq \nbNodes} \abs{\alpha_k}$. \\[-0.15cm]
	
	\begin{itemize}
	
		\item \textbf{Case 1: $\textbf{\textsc{Alg}} = \text{\textsc{Esprit}}$.}
		
		Assume that $\widetilde{\boldsymbol{\Phi}} \triangleq \widetilde{\subspaceMatrix}_\downarrow^\dagger\widetilde{\subspaceMatrix}_\uparrow$ is a solution of the linear system $\widetilde{\subspaceMatrix}_\downarrow\mathbf{Y} = \widetilde{\subspaceMatrix}_\uparrow$ and 
		\begin{equation}
			\gamma \triangleq \frac{\sqrt{\min\{\sizeData, \nbSamples-\sizeData+1\}}\norm{\noiseVector}_2}{\alpha_\mathrm{min}\sigma_\mathrm{min}(\vander_L)\sigma_\mathrm{min}(\vander_{\nbSamples-\sizeData+1})} < \frac{1}{1 + \sqrt{2}\beta}.
			\label{eq: gamma ESPRIT} 
		\end{equation}
		Then, one can find a permutation $\pi$ of $\{1, 2, \ldots, \nbNodes\}$ such that for all $k \fromto{1, 2, \ldots, \nbNodes}$,
		\begin{align}
			\abs{\widehat{\node}_{\pi(k)} - \node_k} \leq \frac{(2\nbNodes-1)\sqrt{2}\beta\gamma}{1 - (1+\sqrt{2}\beta)\gamma}\big(1 + \kappa(\vander_L)\big)\kappa(\vander_L),
			\label{eq: estimation error esprit} 
			\\[-0.7cm]\notag
		\end{align}
		where $\beta \triangleq \frac{\sigma_\mathrm{max}(\vander_\sizeData)}{\sigma_\mathrm{min}(\vander_{\sizeData-1})}$.\\[-0.17cm]
	
		\item \textbf{Case 2: $\textbf{\textsc{Alg}} = \text{\textsc{Mp}}$.}
		
		Assume that the pair $(\widetilde{\boldsymbol{\Psi}}_1, \widetilde{\boldsymbol{\Psi}}_2)$ defined in \eqref{eq: definition psi 1},\eqref{eq: definition psi 2} is regular and
		\begin{equation}
			\gamma \triangleq \frac{\sqrt{\min\{\sizeData, \nbSamples-\sizeData\}}\norm{\noiseVector}_2}{\alpha_\mathrm{min}\sigma_\mathrm{min}(\vander_L)\sigma_\mathrm{min}(\vander_{\nbSamples-\sizeData})} < 1.
			\label{eq: gamma MP method} 
		\end{equation}
		Then, one can find a permutation $\pi$ of $\{1, 2, \ldots, \nbNodes\}$ such that for all $k \fromto{1, 2, \ldots, \nbNodes}$,
		\begin{align}
			\chi(\widehat{\node}_{\pi(k)}, \node_k) &\leq \!\frac{(2\nbNodes-1)\gamma}{\sqrt{1+A_\mathrm{min}^2}}\!\left[\!\frac{2\sqrt{2}}{1 - \gamma} \frac{\alpha_\mathrm{max}}{\alpha_\mathrm{min}}\kappa(\vander_\sizeData)\kappa(\vander_{\nbSamples-\sizeData}) \right. \notag \\
				&\hspace{2cm} \left.+ \left(1 + \frac{\sqrt{2}\gamma}{1 - \gamma}\right)\right],
				\label{eq: perturbation bound MP method chordal metric} 
				\\[-0.6cm]\notag
		\end{align}
		where $A_\mathrm{min} \triangleq \displaystyle \min_{1 \leq k \leq \nbNodes} \abs{\node_k}$.
		
	\end{itemize}
\end{thm}

The proof of Theorem \ref{thm: performance analysis for both MP and LS-ESPRIT} for both the ESPRIT algorithm and the MP method is based on a perturbation result~\cite[p.~102]{Wedin1972} for the singular space of a matrix, which provides us with an upper bound on the principal angle between the noiseless and the noisy signal subspace. 
For the ESPRIT algorithm, we further apply the Bauer-Fike Theorem~\cite[Thm.~IV.3.3]{Stewart1990}, and for the MP method we use a generalization of the Bauer-Fike Theorem to the generalized eigenvalue problem~\cite[Thm.~VI.2.7]{Stewart1990}.

We next turn the bound \eqref{eq: perturbation bound MP method chordal metric} into a bound in terms of Euclidean distance between the estimated nodes $\widehat{\node}_1, \widehat{\node}_2, \ldots, \widehat{\node}_\nbNodes$ and the true nodes $\node_1, \node_2, \ldots, \node_\nbNodes$. 

\begin{cor}
	\label{cor: performance analysis matrix pencil method} 
	Assume that the conditions for Case 2 of Theorem~\ref{thm: performance analysis for both MP and LS-ESPRIT}  are satisfied. Define 
	\begin{align}
		d &\triangleq \frac{(2\nbNodes-1)\gamma}{\sqrt{1+A_\mathrm{min}^2}}\!\left[\!\frac{2\sqrt{2}}{1 - \gamma} \frac{\alpha_\mathrm{max}}{\alpha_\mathrm{min}}\kappa(\vander_\sizeData)\kappa(\vander_{\nbSamples-\sizeData}) \right. \notag\\
				&\hspace{2cm} \left.+ \left(1 + \frac{\sqrt{2}\gamma}{1 - \gamma}\right)\right], \label{eq: d in corollary MP method} 
	\end{align}
	and assume that $d < 1/\sqrt{2}$. 	
	Then, one can find a permutation $\pi$ of $\{1, 2, \ldots, K\}$ such that the estimates $\widehat{\node_k}$, $k \fromto{1, 2, \ldots, \nbNodes}$, delivered by $\text{\textsc{MP}}(\noisySigVector, \nbNodes, \sizeData)$ satisfy
	\begin{equation}
		\abs{\node_k - \widehat{\node}_{\pi(k)}} \leq \eta_k,
		\label{eq: error bound MP method} 
	\end{equation}
	where 
	\begin{equation*}
		\eta_k \triangleq \frac{d\sqrt{1 - d^2}\!\left(1 + \abs{z_k}^2\right)}{1 - d^2\!\left(1 + \abs{z_k}^2\right)} + \left(\!1 - \frac{1}{1 - d^2\!\left(1+\abs{z_k}^2\right)}\!\right)\!\!\abs{z_k},
	\end{equation*}
	for all $k \fromto{1, 2, \ldots, \nbNodes}$.
\end{cor}

This result is derived using \cite[Lem.~7.16]{Anderson1997}, which expresses balls with respect to the chordal distance in terms of Euclidean quantities. We finally note that the condition $d < 1/\sqrt{2}$ is satisfied as long as the noise energy $\norm{\noiseVector}_2$ remains sufficiently small.

Our upper bounds \eqref{eq: estimation error esprit}, \eqref{eq: perturbation bound MP method chordal metric}, and \eqref{eq: error bound MP method} reflect correctly that the estimates $\widehat{\node}_k$ are perfect in the noiseless case. This is seen by noting that $\noiseVector = \boldsymbol{0}$ implies $\gamma = 0$ in \eqref{eq: gamma ESPRIT} and \eqref{eq: gamma MP method} and hence also $d = 0$ in \eqref{eq: d in corollary MP method}.
Theorem~\ref{thm: performance analysis for both MP and LS-ESPRIT} and Corollary~\ref{cor: performance analysis matrix pencil method} show that $\node_1, \node_2, \ldots, \node_\nbNodes$ can be recovered stably (with respect to the dependence of the estimation errors \eqref{eq: estimation error esprit} and \eqref{eq: error bound MP method} on $\norm{\boldsymbol{e}}_{2}$) from the measurements $\noisySig_0, \noisySig_1, \ldots, \noisySig_{\nbSamples-1}$ both via the ESPRIT algorithm and the MP method. 
We emphasize that noise here is deterministic and does not have to satisfy any other condition apart from being small enough so that $\gamma < 1$ and additionally $d < 1/\sqrt{2}$ in the case of the MP method. The condition $d < 1/\sqrt{2}$ guarantees that the estimated nodes $\widehat{\node}_1, \widehat{\node}_2, \ldots, \widehat{\node}_\nbNodes$ do not take on the value $\infty$.
The error bounds on $\abs{\widehat{\node}_k - \node_k}$, for all $k \fromto{1, 2, \ldots, \nbNodes}$, both for the ESPRIT algorithm and the MP method, depend on the minimum and maximum singular values of the Vandermonde matrices $\vander_\sizeData$, $\vander_{\nbSamples-\sizeData}$, $\vander_{\sizeData-1}$, and $\vander_{\nbSamples-\sizeData+1}$. New lower and upper bounds on these quantities, presented in Section~\ref{sec: conditioning Vandermonde matrices}, allow us to express our error estimates in terms of the minimum wrap-around distance between the node frequencies $f_k$, the quantity $\max_{1 \leq k \leq \nbNodes} d_k$, the sampling frequency $F_\mathrm{s}$, and $\nbSamples$, $\sizeData$, $\alpha_\mathrm{min}$, and $\alpha_\mathrm{max}$. Specifically, these results allow us to conclude that the node estimation errors both for the ESPRIT algorithm and the MP method remain small if i) the noise level is small enough, ii) the minimum wrap-around distance between the node frequencies $f_k$ is large relative to $F_\mathrm{s}/(\nbSamples-1)$, and iii) the nodes $z_k$ remain sufficiently close to the unit circle (i.e., the damping factors $d_k$ are sufficiently small). 

In \cite{Hua1988, Hua1990}, Hua and Sarkar employ a first-order perturbation analysis to compare the performance of the MP method to a variant of the Prony method, but this analysis is of statistical nature and requires a high-SNR assumption. The only deterministic result we are aware of for the MP method is due to Moitra~\cite{Moitra2014} who analyzes a new variant of the MP method. Specifically, Moitra replaces the matrices $\dataMatrixOne$ and $\dataMatrixTwo$ in \eqref{eq: expression X1 MP method} and \eqref{eq: expression X2 MP method} by $\mathbf{A} \triangleq \vander_L\mathbf{D}_\alpha\vander_L^H$ and $\mathbf{B} \triangleq \vander_L\mathbf{D}_{\boldsymbol{\alpha}}\mathbf{D}_{\boldsymbol{z}}\vander_L^H$, respectively. 
The corresponding results apply to undamped sinusoids, i.e., $\abs{\node_k} = 1$, $k \fromto{1, 2, \ldots, \nbNodes}$, only. Moitra's proof technique reveals an interesting connection between the condition number of Vandermonde matrices with nodes on the unit circle and Selberg's work on the large sieve inequality~\cite{Selberg1991}. 

In \cite{Eriksson1993} and \cite{Stoica1991}, it is shown, for undamped sinusoids, that the ESPRIT algorithm has asymptotic ($\mathrm{SNR} \rightarrow \infty$ in \cite{Eriksson1993} and $N \rightarrow \infty$ in \cite{Stoica1991}) statistical efficiency close to $1$.  
In \cite{Rao1988}, in the context of direction of arrival estimation, expressions for the asymptotic mean squared error are derived for ESPRIT for the undamped case under a high--$\mathrm{SNR}$ assumption. In \cite{Hua1991}, it is shown that in the case of undamped sinusoids, the ESPRIT algorithm and the MP method are less sensitive to noise than MUSIC. A unified performance analysis for $\mathrm{SNR} \rightarrow \infty$ that applies to both the ESPRIT algorithm and the MP method is proposed in \cite{Li1991}. All these performance analyses are of statistical and asymptotic nature. We are not aware of any non-asymptotic and deterministic performance analyses for the ESPRIT algorithm, like the one performed here.


\section{New bounds on the minimum and maximum singular values of Vandermonde matrices}
\label{sec: conditioning Vandermonde matrices} 
 
 In this section, we provide new lower and upper bounds on the minimum and maximum singular values of Vandermonde matrices with nodes inside the unit disk. 
In order to put our results into perspective, we first review bounds available in the literature.
An upper bound on the condition number of Vandermonde matrices with nodes inside the unit disk was provided by Baz\'an in~\cite[Thm.~6]{Bazan2000}. This bound is, however, somewhat complicated and seems to be amenable to analytical statements only for $N \rightarrow \infty$. Specifically, it allows to conclude that the condition number is close to $1$ if the nodes are separated enough and close to the unit circle. Unlike Baz\'an's result~\cite[Thm.~6]{Bazan2000}, the upper bound on the condition number we present here is expressed directly in terms of the minimum distance of the nodes from the unit circle.
Our result is inspired by the link---first established by Moitra~\cite{Moitra2014}---between the condition number of Vandermonde matrices with nodes on the unit circle and Selberg's work on sharp forms of the large sieve inequality~\cite{Selberg1991}.
We rely on a result by Montgomery and Vaaler~\cite{Montgomery1998} extending---to the complex case---a generalization of Hilbert's inequality due to Montgomery and Vaughan~\cite[Thm.~1]{Montgomery1974}. 
In contrast, the derivation of Moitra's upper bound is based on extremal minorants and majorants for the characteristic function of an interval. Both Moitra's result and our result are, however, in essence, linked to the large sieve inequality.

\begin{thm}
	\label{thm: upper bound spectral condition number nodes in the unit disk refinement} 
	For $k \fromto{1, 2, \ldots, \nbNodes}$, let $z_k \triangleq e^{-d_\mathrm{max}}e^{2\pi if_k/F_\mathrm{s}}$ be complex numbers with $d_k \geq 0$ and $f_k \in [0, F_\mathrm{s})$. 
	Let 
	\begin{equation*}
		\delta \triangleq \min_{n \in \Z} \min_{\substack{1 \leq k, \ell \leq K \\ k \neq \ell}} \abs{f_k - f_\ell + nF_\mathrm{s}}
	\end{equation*}
	be the minimum wrap-around distance between the $f_k$, $k \fromto{1, 2, \ldots, \nbNodes}$,
	and $d_\mathrm{max} \triangleq \displaystyle \max_{1 \leq k \leq \nbNodes} d_k$.
	For 
	\begin{equation}
		d_\mathrm{max} < 1/(\nbSamples-1)
		\label{eq: condition on dmax} 
	\end{equation}
	and
	\begin{equation}
		\delta > \frac{84F_\mathrm{s}}{\pi\left(\nbSamples-1\right)\big(1 - d_\mathrm{max}(\nbSamples-1)\big)},
		\label{eq: condition on delta} 
	\end{equation}
	the smallest and largest singular values of the Vandermonde matrix $\vander_N$ obey
	\begin{align*}
		\sigma_\mathrm{min}^2(\vander_N) &\geq \left(\nbSamples-1\right)\big(1 - d_\mathrm{max}(\nbSamples-1)\big) - 84F_\mathrm{s}/(\pi\delta)\\
		\sigma_\mathrm{max}^2(\vander_N) &\leq \nbSamples-1 + 84F_\mathrm{s}/(\pi\delta),
	\end{align*}
	and thus, the condition number of $\vander_N$ satisfies
	\begin{equation}
		\kappa\!\left(\vander_\nbSamples\right) \leq \sqrt{\frac{\nbSamples-1 + 84F_\mathrm{s}/(\pi\delta)}{\left(\nbSamples-1\right)\big(1 - d_\mathrm{max}(\nbSamples-1)\big) - 84F_\mathrm{s}/(\pi\delta)}}.
		\label{eq: main result upper bound spectral condition number for nodes in the unit disk} 
	\end{equation}
\end{thm}
	
Theorem~\ref{thm: upper bound spectral condition number nodes in the unit disk refinement} shows that the condition number of Vandermonde matrices with nodes in the unit disk is close to $1$ if the minimum wrap-around distance between the node frequencies $f_k$ is large relative to $F_\mathrm{s}/(\nbSamples-1)$, and the damping factors $d_k$ are small enough (i.e., the nodes $z_k$ are close enough to the unit circle). The conditions \eqref{eq: condition on dmax} and \eqref{eq: condition on delta} on $d_\mathrm{max}$ and $\delta$ ensure that our lower bound on $\sigma_\mathrm{min}(\vander_\nbSamples)$ is positive.
When particularized for the undamped case $d_\mathrm{max} = 0$ (i.e., $\abs{\node_k} = 1$ for all $k \fromto{1, 2, \ldots, \nbNodes}$), our result recovers Moitra's upper bound provided in \cite[Thm.~2.3]{Moitra2014} up to a difference in the constant $84/\pi$ in the numerator and denominator of \eqref{eq: main result upper bound spectral condition number for nodes in the unit disk}, which in Moitra's case ($d_\mathrm{max} = 0$) equals $1$. We note, however, that for $d_\mathrm{max} = 0$, \cite[Thm.~1]{Montgomery1974} can be used instead of \cite{Montgomery1998} to recover Moitra's upper bound exactly in our approach.



%
\bibliographystyle{IEEEtran} 
\bibliography{ref}

\end{document}